\documentstyle[12pt]{article}
\def\ie{{\em i.e.}}
\def\eg{{\em e.g.}}

\def\beq{\begin{equation}}
\def\eeq{\end{equation}}

\def\Tr#1{{\rm Tr}\,#1}
\catcode`\@=11 % This allows us to modify PLAIN macros.
\def\coeff#1#2{{\textstyle{#1\over #2}}}

\def\vev#1{\langle #1\rangle}
\def\lsim{\mathrel{\mathpalette\@versim<}}
\def\gsim{\mathrel{\mathpalette\@versim>}}
\def\@versim#1#2{\vcenter{\offinterlineskip
    \ialign{$\m@th#1\hfil##\hfil$\crcr#2\crcr\sim\crcr } }}
\def\etal{{\em et. al.}}
\def\JL{J. L. Lopez}
\def\DVN{D. V. Nanopoulos}
\def\AZ{A. Zichichi}

\def\r#1{$\bf#1$}
\def\rb#1{$\bf\overline{#1}$}

\def\t1{{\tilde 1}}

\def\eV{\,{\rm eV}}

\def\GeV{\,{\rm GeV}}
\def\TeV{\,{\rm TeV}}

\def\y{\,{\rm y}}

\def\lra{\leftrightarrow}
\def\to{\rightarrow}

\def\NPB#1#2#3{Nucl. Phys. B {\bf#1} (19#2) #3}
\def\PLB#1#2#3{Phys. Lett. B {\bf#1} (19#2) #3}

\def\PRD#1#2#3{Phys. Rev. D {\bf#1} (19#2) #3}
\def\PRL#1#2#3{Phys. Rev. Lett. {\bf#1} (19#2) #3}

\def\MODA#1#2#3{Mod. Phys. Lett. A {\bf#1} (19#2) #3}
\def\IJMP#1#2#3{Int. J. Mod. Phys. A {\bf#1} (19#2) #3}

\def\hepph#1{({\tt hep-ph/#1})}
\def\hepth#1{({\tt hep-th/#1})}

\begin{document}
\begin{flushright}
\baselineskip=12pt
{CTP-TAMU-29/95}\\
{ACT-11/95}\\
{\tt hep-ph/9508253}
\end{flushright}

\begin{center}
\vglue 0.5cm
{\Large\bf The Missing Doublet Model Revamped\\}
\vglue 0.5cm
{\Large Jorge L. Lopez$^{1,2,\dagger}$ and D.V. Nanopoulos$^{1,2}$\\}
\vglue 0.25cm
\begin{flushleft}
$^1$Center for Theoretical Physics, Department of Physics, Texas A\&M
University\\ College Station, TX 77843--4242, USA\\
$^2$Astroparticle Physics Group, Houston Advanced Research Center (HARC)\\
The Mitchell Campus, The Woodlands, TX 77381, USA\\
\end{flushleft}
\end{center}

\vglue 0.5cm
\begin{abstract}
We revisit the Missing Doublet Model (MDM) as a means to address the
apparent difficulties of the minimal $SU(5)$ supergravity model in dealing
with the doublet-triplet splitting problem, the prediction of $\alpha_3(M_Z)$,
and the proton lifetime. We revamp the original MDM by extending its
observable sector to include fields and interactions that naturally suppress
the dimension-five proton decay operators and that allow see-saw neutrino
masses. We also endow the model with a hidden sector which (via gaugino
condensation) triggers supersymmetry breaking of the desired magnitude, and
(via hidden matter condensation) yields a new dynamical intermediate scale
for the right-handed neutrino masses ($\sim10^{10}\GeV$), and provides an
effective Higgs mixing parameter $\mu$. The model is consistent with gauge
coupling unification for experimentally acceptable values of $\alpha_3(M_Z)$,
and with proton decay limits even for large values of $\tan\beta$. The
right-handed neutrinos can be produced subsequent to inflation, and their
out-of-equilibrium decays induce a lepton asymmetry which is later reprocessed
(via sphaleron interactions) into a baryon asymmetry at the electroweak scale.
The resulting see-saw neutrino masses provide a candidate for the hot dark
matter component of the Universe ($m_{\nu_\tau}\sim{\cal O}(10\eV)$) and are
consistent with the MSW solution to the solar neutrino problem. We finally
compare the features of this traditional GUT model with that of the
readily string-derivable $SU(5)\times U(1)$ model, and discuss the prospects
of deriving the revamped MDM from string theory.
\end{abstract}
\vspace{0.1cm}
\begin{flushleft}
\baselineskip=12pt
{CTP-TAMU-29/95}\\
{ACT-11/95}\\
August 1995
\end{flushleft}
\bigskip
\hrule
\smallskip
{\footnotesize $^\dagger$ Address after August 15, 1995: Department of Physics,
Rice University, Houston, TX 77005, USA.}

\vfill\eject
\setcounter{page}{1}
\pagestyle{plain}
\baselineskip=14pt

\section{Introduction}
\label{sec:Introduction}
The much heralded convergence of the Standard Model gauge couplings in
supersymmetric Grand Unified Theories (GUTs) \cite{EKN}, is continually being
challenged by ever more precise LEP measurements of the gauge couplings. It was
realized early on that the effect of the GUT particles responsible for the
onset of the unified theory, was not negligible \cite{GUTthresholds}. However,
because of the presumed great uncertainty in the GUT physics, such discussions
have been largely carried out in the context of the minimal $SU(5)$
supergravity model \cite{CAN}. Central to the study of these issues is the
technical point of how exactly these GUT (or lighter) particles decouple at
scales below their masses. Recent investigations \cite{smooth} reveal that a
``smooth" decoupling of particles leads to noticeable differences from the
step-function approximation. Moreover, these new effects coupled with the
latest LEP data on $\sin^2\theta_W$ and the determination of the top-quark
mass, lead to a greatly increased prediction for $\alpha_3(M_Z)$
\cite{Clavelli,Bagger,LP}, strongly suggesting that minimal $SU(5)$ GUT
thresholds are unable to bring the $\alpha_3(M_Z)$ prediction down to the
experimentally acceptable range \cite{Clavelli}. This impasse may be resolved
with a significant contribution from Planck-scale non-renormalizable operators
\cite{NROs,LP}, although such effects call into question the whole
field-theoretical approximation to the gauge coupling unification problem.

Even if Planck-scale physics can resolve the present $\alpha_3$ discrepancy
in minimal $SU(5)$, this GUT model suffers from a well known fine-tuning
\cite{DG} regarding the solution of the doublet-triplet splitting problem of
the Higgs pentaplets. At least three solutions to this problem (all involving
non-minimal GUT models) have been proposed: the missing-partner mechanism
\cite{MPM}, the sliding-singlet mechanism \cite{SSM}, and the
pseudo-goldstone-boson mechanism \cite{PGBM}. In the sliding-singlet mechanism
radiative corrections destroy the gauge hierarchy \cite{Lahanas}, whereas an
additional global or local $SU(6)$ symmetry is required in the
pseudo-goldstone-boson mechanism. It is very suggestive that the same
investigations that uncover the $\alpha_3$ discrepancy in minimal $SU(5)$, also
show that in the so-called Missing Doublet Model (MDM) \cite{MPM}, which has as
its central component the missing-partner mechanism, the $\alpha_3$ prediction
is decreased to acceptable values \cite{Bagger,Clavelli}. As we discuss below,
some variants of the missing-doublet model \cite{MNTY2,HMTY} also solve the
problematic situation with dimension-five proton decay operators in minimal
$SU(5)$, which require the Higgs triplet mass ($M_{H_3}$) to exceed the GUT
scale and the supersymmetric spectrum to be tuned in specific ways
\cite{AN,HMY}, especially when cosmological constraints are simultaneously
enforced \cite{LNP}. In fact, an updated analysis has recently shown
\cite{HMTYpd} that the upper bound on the Higgs triplet mass from unification
constraints (\ie,
$M_{H_3}\propto e^{-5\pi/3\alpha_3},\ \alpha_3<\alpha_3^{\rm max}$), and the
corresponding lower bound from proton decay constraints (\ie, $\tau_p\propto
M^2_{H_3},\ \tau_p>\tau^{\rm min}_p$) are very close to each
other, leaving only a small window of allowed parameter space in minimal
$SU(5)$. Note also that, because of the rather large representations introduced
in the MDM (\r{75},\r{50},\rb{50}), it is necessary to assign some of these
Planck-size masses, in order to avoid the onset of a strongly-interacting GUT
below the Planck scale \cite{Clavelli,HMTY}. Thus, Planck-scale physics is
again unavoidable in this more realistic version of $SU(5)$ GUTs.

$SO(10)$ GUTs \cite{GFM,GN} have also received a great deal of attention lately
\cite{BB,Mohapatra,Raby,inspired}, with interesting successes in the area of
quark and lepton masses and mixings, although the $\tan\beta={\cal O}(50)$
prediction requires fine-tuning of the supersymmetric spectrum \cite{NR,Carena}
to reconcile it with radiative electroweak breaking. Assuming universal soft
supersymmetry breaking, the further constraints from $B(b\to s\gamma)$ and
cosmology strongly disfavor the model \cite{BOP}. However, most of these
shortcomings are overcome when one allows certain classes of non-universal
scalar masses \cite{OP,BOP}. More to the point, the successes of $SO(10)$ rely
on the existence of certain non-renormalizable operators (as originally
suggested in Ref.~\cite{EG+NS}) that are presumed to be obtained from a
string-derived model at the Planck scale. Despite initial claims \cite{Lykken},
no {\em consistent} $SO(10)$ GUT string model has been derived in the context
of free-fermionic strings \cite{Cleaver}. However, these failed attempts have
been enough to fuel a series of ``string-inspired" $SO(10)$ GUT models
\cite{inspired}, which are limited to certain type and number of
representations (those allowed by the level-two Kac-Moody
construction\footnote{At level two, the allowed unitary massless
representations are \r{1},\r{10},\r{16},\rb{16},\r{45},\r{54} \cite{ELN}. A
string model containing the \r{126},\rb{126} representations used in
traditional $SO(10)$ model building requires an unlikely level-five
construction \cite{ELN}.}), forcing model builders to rely heavily on
postulated effective non-renormalizable operators \cite{inspired}. Level-two
string $SO(10)$ GUT models have been consistently constructed in the context of
symmetric orbifolds \cite{Aldazabal}, but with limited phenomenological
success, especially in dealing with the doublet-triplet splitting problem.

In view of its field-theoretical successes, in this paper we revisit the
missing-doublet model as a well-motivated, realistic contender for a grand
unified model. We first review the original MDM and its features and
shortcomings (Sec.~\ref{sec:observable}). We then propose a simple extension of
the model to naturally suppress dimension-five proton decay operators
(Sec.~\ref{sec:observable}). Our most substantive contribution is to endow this
{\em supergravity} model with a hidden sector containing gauge and matter
degrees of freedom (Sec.~\ref{sec:hidden}). Hidden sector gaugino condensation
triggers supersymmetry breaking which, as we discuss, can be of the desired
magnitude for suitable choices of the hidden gauge group and hidden matter
spectrum. The matter condensates provide a new dynamical intermediate scale
which, via non-renormalizable interactions, generates a low-energy Higgs mixing
term $\mu$. With the introduction of right-handed neutrinos to the model, this
scale also becomes their mass scale, which provides a suitable see-saw spectrum
of neutrino masses (Secs.~\ref{sec:observable},\ref{sec:CBA}). We show that the
model is consistent with gauge coupling unification for experimentally
acceptable values of $\alpha_3(M_Z)$ and that dimension-five proton decay
operators are consistent with present limits even for large values of
$\tan\beta$ (Sec.~\ref{sec:unification}). Also, the out-of-equilibrium decays
of the right-handed neutrinos subsequent to inflation produce a lepton
asymmetry which is re-processed into a baryon asymmetry by strongly-interacting
Standard Model effects (sphalerons) at the electroweak scale
(Sec.~\ref{sec:CBA}).  We finally compare the features of this traditional GUT
model with that of the readily string-derivable $SU(5)\times U(1)$ model, and
discuss the prospects of deriving the revamped MDM from string theory
(Sec.~\ref{sec:comparison}). We summarize our conclusions in
Sec.~\ref{sec:conclusions}.

\section{The observable sector}
\label{sec:observable}
The original MDM \cite{MPM} can be described by the following set of
observable sector fields: $\Sigma$ (\r{75}), $\theta$ (\r{50}), $\bar\theta$
(\rb{50}), $h$ (\r{5}), $\bar h$ (\rb{5}), $F_{1,2,3}$ (\r{10}'s),
$\bar f_{1,2,3}$ (\rb{5}'s), interacting via the superpotential
\begin{equation}
W=\coeff{1}{2}M_{75}\,\Tr{\Sigma^2}+\coeff{1}{3}\lambda_{75}\,\Tr{\Sigma^3}
+\lambda_4\,\bar\theta\Sigma h+\lambda_5\,\theta\Sigma\bar h
+M_{50}\,\theta\bar\theta+\lambda^{ij}_2 F_iF_jh
+\lambda^{ij}_1 F_i\bar f_j \bar h\ .
\label{eq:W}
\end{equation}
The expectation value of the \r{75} can be chosen such that the $SU(5)$ gauge
symmetry is broken down to the Standard Model one, in which case the scalar
potential that follows from Eq.~(\ref{eq:W}) gives
$\vev{\Sigma}\sim M_{75}/\lambda_{75}$. The $\bar\theta\Sigma h$,
$\theta\Sigma\bar h$, and $\theta\bar\theta$ terms in $W$ effect the
doublet-triplet mechanism via the mass matrix
\begin{equation}
\bordermatrix{
&\bar h_3&\bar\theta_3\cr
h_3&0&\lambda_4\vev{\Sigma}\cr
\theta_3&\lambda_5\vev{\Sigma}&M_{50}}\ ,
\label{eq:2/3}
\end{equation}
where the subscript `3' indicates the $SU(2)_L$ singlet, $SU(3)_C$ triplet
component of the corresponding $SU(5)$ representation. This matrix clearly
yields massive ($\sim\vev{\Sigma}\sim M_{\rm GUT}$) Higgs triplets ($h_3,\bar
h_3$), whereas the doublets ($h_2,\bar h_2$) remain massless. The $M_{50}$ term
is not required for a successful doublet-triplet splitting. However, it is
introduced in order to give large masses to the many leftover components of the
\r{50},\rb{50} representations. The last two terms in W (\ref{eq:W}) provide
the Yukawa matrices for the Standard Model fermions, implying the usual
relations (\eg, $\lambda_b=\lambda_\tau$).

Despite the above natural solution to the doublet-triplet splitting problem,
the magnitude of the dimension-five ($d=5$) proton decay operators still needs
to be assessed. The crucial element in this calculation is the effective
$h_3\bar h_3$ mixing term. If in Eq.~(\ref{eq:W}) $M_{50}$ were allowed to
vanish, then there would be no mixing whatever, and the $d=5$ operators would
be negligible. In practice $M_{50}$ cannot vanish, and we are left with two
possibilities: (i) $M_{50}\sim \vev{\Sigma}$, and (ii)
$M_{50}\gg\vev{\Sigma}$. The first case implies an effective Higgs-triplet
mixing term of the same magnitude as in the minimal $SU(5)$ model, and
therefore similar difficulties in suppressing proton decay. However, this case
is not really viable since above the GUT scale the large \r{50},\rb{50}
representations increase the $SU(5)$ beta function so much that the gauge
coupling becomes non-perturbative before reaching the Planck scale
\cite{Clavelli,HMTY}. We are left with the second alternative with $M_{50}\sim
M$, where $M=M_{Pl}/\sqrt{8\pi}\approx10^{18}\GeV$ is the appropriate
gravitational scale. Unfortunately, this choice leads to a see-saw type mass
for the Higgs triplets: $m_{h_3,\bar h_3}\sim \vev{\Sigma}^2/M\sim10^{14}\GeV$,
and effective mixing of the same magnitude, which makes proton decay much too
fast.

Various variants of the MDM have been proposed to deal with the proton decay
problem in a more effective way \cite{MPM,MNTY2,HMTY}. Here we follow the
suggestion in Ref.~\cite{HMTY}, whereby the following additional fields are
introduced: $\theta'$ (\r{50}), $\bar\theta'$ (\rb{50}), $h'$ (\r{5}), $\bar
h'$ (\rb{5}). The superpotential for the model is that in Eq.~(\ref{eq:W}) with
$M_{50}\equiv0$, and supplemented by
\begin{equation}
W'=\lambda_4'\,\bar\theta'\Sigma h'+\lambda_5'\,\theta'\Sigma\bar h'
+M'_{50}\,\theta\bar\theta'+M'_{50}\,\theta'\bar\theta\ ,
\label{eq:W'}
\end{equation}
where we again take $M'_{50}\sim M$. These interactions lead to the following
generalized Higgs-triplet mass matrix
\begin{equation}
\bordermatrix{
&\bar h'_3&\bar\theta_3&\bar h_3&\bar\theta'_3\cr
h_3&0&\lambda_4\vev{\Sigma}&0&0\cr
\theta'_3&\lambda'_5\vev{\Sigma}&M'_{50}&0&0\cr
h'_3&0&0&0&\lambda'_4\vev{\Sigma}\cr
\theta_3&0&0&\lambda_5\vev{\Sigma}&M'_{50}}\ ,
\label{eq:2/3'}
\end{equation}
and effective interactions \cite{HMTY}
\begin{equation}
\left(\lambda_4\lambda'_5{\vev{\Sigma}^2\over M'_{50}}\right)h_3\bar h'_3+
\left(\lambda'_4\lambda_5{\vev{\Sigma}^2\over M'_{50}}\right)h'_3\bar h_3
\equiv M_{H_3}\,h_3\bar h'_3+M_{\bar H_3}\,h'_3\bar h_3\ .
\label{eq:mixings}
\end{equation}
Since there is no effective interaction between $h_3$ and $\bar h_3$ (the
only triplets that interact with the Standard Model fermions), the $d=5$
operator is negligible.

If the superpotential $W+W'$ were the complete model, we would have managed
to make all the non-minimal fields sufficiently heavy or non-interacting.
However, we would have left two pairs of Higgs doublets $h_2,\bar h_2$ and
$h'_2,\bar h_2'$ with no apparent use for the second pair, and if light, with
severe trouble with gauge coupling unification. Let us assume the existence of
a mass term $M_{h'}h'\bar h'$, with no specific origin for $M_{h'}$ for now.
Such a term contains $M_{h'}h'_3\bar h'_3$, which ``hooks up" the two
disconnected pieces in Eq.~(\ref{eq:mixings}) and allows $d=5$ proton decay to
occur, with an operator proportional to
\begin{equation}
{1\over M_{H_{\rm eff}}}\equiv{M_{h'}\over M_{H_3} M_{\bar H_3}}
\sim {M_{h'}\over [\vev{\Sigma}^2/M'_{50}]^2}\ ,
\label{eq:Heff}
\end{equation}
where $M_{H_{\rm eff}}$ is the effective Higgs triplet mass. Since in the
minimal $SU(5)$ model with $M_{H_{\rm eff}}=M_{H_3}\gsim 10^{17}\GeV$, the
present experimental bounds on proton decay are satisfied without strong
restrictions on the parameter space \cite{AN,HMY}, we effectively require
$M_{h'}\lsim10^{11}\GeV$.

But where does this intermediate scale come from? It has been suggested that
this scale could be generated dynamically via the breaking of a Peccei-Quinn
symmetry \cite{MSY,HMTY}. A more modern and economical approach to the
generation of intermediate scales, especially in the context of supergravity,
is to consider the condensation of a hidden sector gauge group that triggers
supersymmetry breaking. Non-renormalizable interactions coupling hidden sector
matter fields to observable fields may then naturally generate the
intermediate scale.\footnote{This mechanism is commonly available in string
model building \cite{decisive}.} Specifically, we add to our model
the following superpotential terms\footnote{The apparent asymmetry between
the $h\bar h$ and $h'\bar h'$ couplings may be understood on the basis
of additional local $U(1)$ quantum numbers, which are broken near the Planck
scale and are carried by both hidden and observable sector particles, as is
common in string model building \cite{decisive}. For further symmetry
arguments motivating these choices, see \eg, Ref.~\cite{CKN}.}
\begin{equation}
W''= \lambda_7\, h\bar h\,{(T\bar T)^2\over M^3}
+\lambda'_7\, h'\bar h'\,{T\bar T\over M}\ ,
\label{eq:W''}
\end{equation}
where $T\bar T$ is a gauge-singlet hidden-sector composite (\eg, \r{4}\rb{4} in
$SU(4)$). When the hidden sector condenses, we generate dynamically two mass
scales:
\begin{equation}
M_{h'}=\lambda'_7\,{\vev{T\bar T}\over M}\,,\qquad
\mu=\lambda_7 \,{\vev{T\bar T}^2\over M^3}\ .
\label{eq:scales}
\end{equation}
Note that for $\vev{T\bar T}/M\sim10^{10}\GeV$, we would obtain for the masses
of the extra pair of doublets $M_{h'}\sim10^{10}\GeV$, and an effective
Higgs-triplet mixing which satisfies proton decay constraints automatically.
We would also obtain dynamically\footnote{This dynamical generation of the
$\mu$ parameter via non-renormalizable interactions is also familiar from
string model-building \cite{decisive,Casasmu,CKN}.} a very desirable Higgs
mixing parameter $\mu\sim100\GeV$. In the next section we explore the hidden
sector of the model with these phenomenological constraints in mind.

One of the main model-building shortcomings of $SU(5)$ GUTs is the
not-so-obvious source of neutrino masses. In fact, neutrino masses can be
introduced by simply adding right-handed neutrino ($SU(5)$ singlet) fields
to the model. To implement the standard see-saw mechanism we introduce
three singlet fields $\nu^c_{1,2,3}$ with the following superpotential
\begin{equation}
W'''=\lambda^{ij}_3\, \bar f_i \nu^c_j h + \lambda^{ij}_6\, \nu^c_i\nu^c_j \,
{T\bar T\over M}\ .
\label{eq:W'''}
\end{equation}
After hidden sector condensation and electroweak symmetry breaking, we
obtain the following see-saw neutrino mass matrix
\begin{equation}
\bordermatrix{&\nu_j&\nu^c_j\cr
\nu_i&0&\lambda_3^{ij}v_2\cr
\nu^c_i&\lambda_3^{ji}v_2&\lambda^{ij}_6\vev{T\bar T}/M}\ ,
\label{eq:see-saw}
\end{equation}
and light neutrino see-saw masses
$m_{\nu}\sim\lambda^3_2 v^2_2/[\lambda_6\vev{T\bar T}/M]$. For simplicity,
in what follows we assume $\lambda^{ij}_6=\lambda^i_6\delta_{ij}$.
With our above desired value of $\vev{T\bar T}/M\sim M_{\nu^c}\sim10^{10}\GeV$,
and $\lambda_3 v_2\sim10\GeV$, typical see-saw light neutrino masses follow,
\ie, $m_{\nu_\tau}\sim10\eV$. Further discussion of the consequences of this
see-saw matrix for the light neutrino masses and mixing angles is given in
Sec.~\ref{sec:CBA} below.

\section{The hidden sector}
\label{sec:hidden}
Our supergravity model is endowed with a hidden sector which communicates
with the observable sector via gravitational interactions (or via $U(1)$
gauge interactions broken near the Planck scale). The hidden sector consists
of a hidden gauge group and a set of matter representations, which for
convenience we take to be $SU(N_c)$ with $N_f$ flavors ($T_i,\bar T_i,\ i=1\to
N_f$) and $N_f<N_c$. This gauge group starts with a gauge coupling $g$ at the
Planck scale ($Q=M$), and becomes strongly interacting at the condensation
scale defined by
\begin{equation}
\Lambda=M e^{8\pi^2/\beta g^2}\ ,
\label{eq:Lambda}
\end{equation}
where the beta function is given by $\beta=-3N_c+N_f$. For simplicity we assume
that all the flavors are ``light", \ie, they have masses\footnote{Massless
flavors lead to pathologies (\ie, no vacuum), which can nonetheless be remedied
by invoking supersymmetry-breaking masses for the $T_i,\bar T_i$ fields
\cite{Peskin}.} $m\ll\Lambda$. At the condensation scale, the strongly
interacting theory is described in terms of composite ``meson" fields $T_i\bar
T_i$. The dynamics of this system can be obtained from an effective Lagrangian
with the following non-perturbative superpotential \cite{Seiberg}
\begin{equation}
W_{\rm non-pert}=(N_c-N_f)\,{\Lambda^{(3N_c-N_f)/(N_c-N_f)}\over
(\det T\bar T)^{1/(N_c-N_f)}}+\Tr(mT\bar T)\ .
\label{eq:Wnp}
\end{equation}
Minimization of the corresponding scalar potential results in the following
expectation values for the mesons fields $\vev{T\bar T}$ (we work in a diagonal
flavor basis)
\begin{equation}
\vev{T\bar T}=\Lambda^{(3N_c-N_f)/N_c}\,(\det m)^{1/N_c}\, m^{-1}
=\Lambda^3 \left({m\over\Lambda}\right)^{N_f/N_c}{1\over m}
=\Lambda^2\, x^{(N_f/N_c)-1}\ ,
\label{eq:TT}
\end{equation}
where in the last expression we have defined $x\equiv m/\Lambda$, with $x<1$.
Inserting this expectation value in $W_{\rm non-pert}$, we obtain
\begin{equation}
\vev{W}=N_c\, \Lambda^3\, x^{N_f/N_c}\ ,
\label{eq:Wvev}
\end{equation}
where $W$ includes all perturbative and non-perturbative contributions. In a
supergravity theory, the scale of supersymmetry breaking is determined by
the gravitino mass: $m_{3/2}=\vev{e^K\,W}$, where $K$ is the K\"ahler
potential. In simple models $K=\sum\phi_i\phi_i^\dagger$, and thus $\vev{K}=0$.
More complicated forms of $K$ are obtained in string models (where the dilaton
and moduli fields play an important role). For our present purposes, we simply
assume that $\vev{e^K}\sim1$. This implies that $\vev{W}$ is the sole source
of supersymmetry breaking, \ie,
\begin{equation}
m_{3/2}\sim \vev{W}\sim \left({\Lambda\over M}\right)^3\, x^{N_f/N_c}\, M\ ,
\label{eq:m3/2}
\end{equation}
where we have restored the units in the expression for $m_{3/2}$.

\begin{figure}[p]
\vspace{6.5in}
\includegraphics{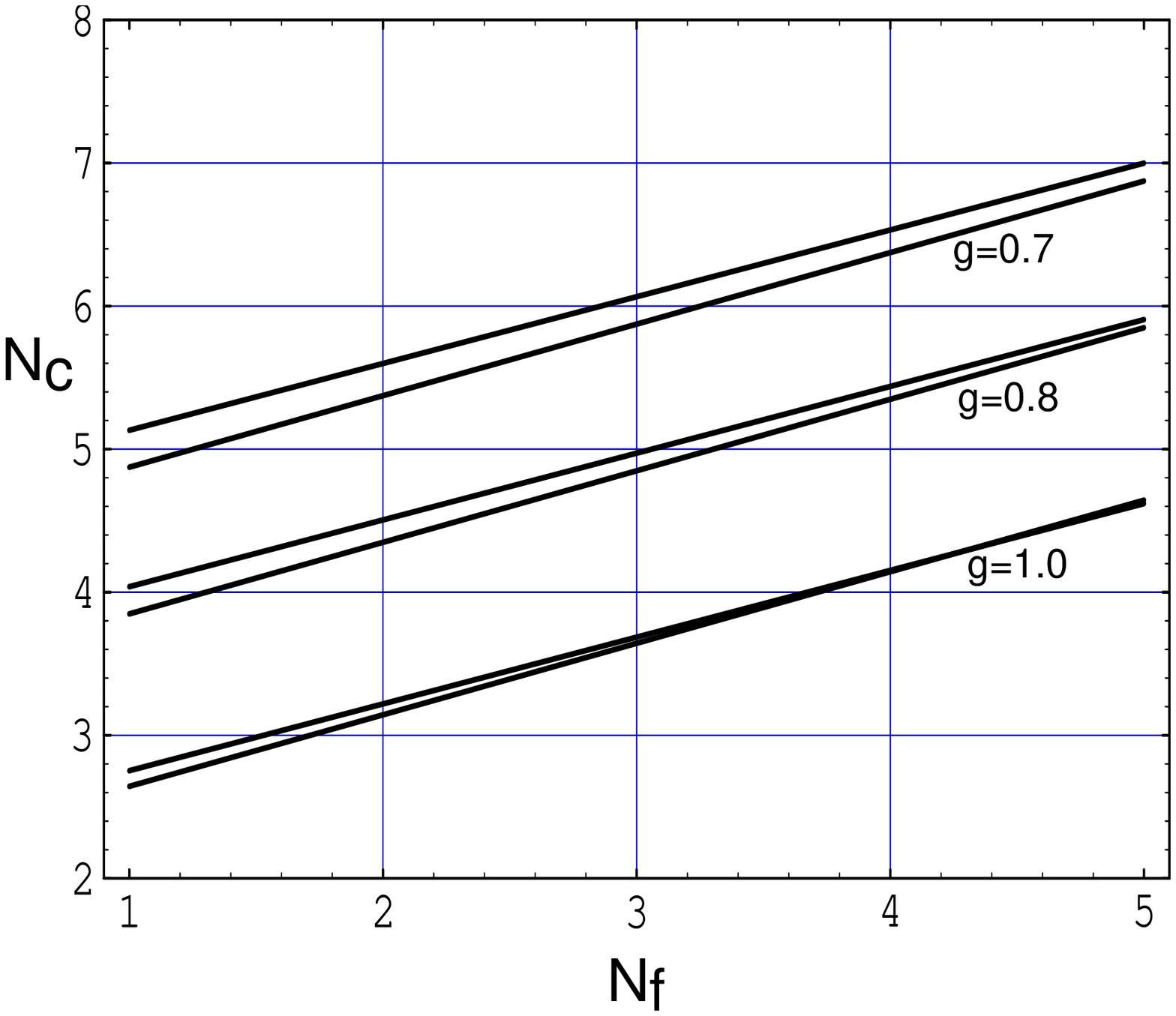}
\caption{Constraints on hidden sector gauge group $SU(N_c)$ with $N_f$ light
flavors, such that the supersymmetry breaking scale ($m_{3/2}$) is between
100 GeV (bottom plots) and 1 TeV (top plots), and the hidden matter condensate
scale is $\vev{T\bar T}/M=10^{10}\,{\rm GeV}$, for different values of the
gauge coupling ($g$) at the Planck scale.}
\label{fig:NcNf}
\end{figure}

With the results in Eqs.~(\ref{eq:TT}) and (\ref{eq:m3/2}) for $\vev{T\bar T}$
and $m_{3/2}$, we can now investigate the conditions on $N_c$, $N_f$, and $x$
that would yield the desired results: $\vev{T\bar T}/M=10^p\GeV$ and
$m_{3/2}=10^q\GeV$, with $p\sim10$ and $q\sim3$. In terms of $p$ and $q$, we
can solve simultaneously Eqs.~(\ref{eq:TT}), (\ref{eq:m3/2}), and
(\ref{eq:Lambda}), to obtain
\begin{equation}
N_c={p-q\over18-q}\,N_f+{8\pi^2\over g^2}\,\,{\log_{10} e\over18-q}\ ,
\label{eq:Nc}
\end{equation}
and
\begin{equation}
x=10^{2N_c\,({3\over2}p-q-9)/\beta}\ .
\label{eq:x}
\end{equation}
Thus, for a given value of $g$ and $N_f$, we obtain $N_c$ (and thus $\beta$)
from Eq.~(\ref{eq:Nc}). With this value of $N_c$, $x$ is determined from
Eq.~(\ref{eq:x}), and $\Lambda$ from Eq.~(\ref{eq:Lambda}). For the desired
$p=10$ and $q=3$, and with the sensible inputs $N_f=1$ and $g=0.7$, we obtain
$N_c=5$, $x\approx0.01$, and $\Lambda\approx10^{13}\GeV$. That is, an $SU(5)$
hidden gauge group with one light flavor. The general constraints on $N_c$ and
$N_f$ for given values of $g$ are shown in Fig.~\ref{fig:NcNf}, for $q=2\to3$
(\ie, $m_{3/2}=100\GeV\to1\TeV$) and $p=10$.

We do not address here the calculation of the observable-sector
soft-supersymmetry-breaking scalar and gaugino masses, since these
depend on the specific choices for the K\"ahler function and the gauge
kinetic function, although their overall scale is already determined by
$m_{3/2}$. The ``flat" choice $K=\sum\phi_i\phi^\dagger_j$ leads to the usual
universal scalar masses, but this choice is not unique.

\section{Unification and proton decay}
\label{sec:unification}
The revamped MDM presented in the two previous sections contains several
departures from conventional gauge coupling unification: (i) there is a pair
of Higgs doublets with intermediate-scale masses ($M_{h'}\sim10^{10}\GeV$),
(ii) there is a richer structure of GUT particles, including two pairs of
Higgs triplets (from the \r{5},\rb{5} representations) with masses
$M_{H_3,\bar H_3}\sim10^{14}\GeV$, and (iii) there is a spectrum of masses for
the different components of the \r{75}, all near the unification scale. There
is also a
hidden gauge group, with an in-principle independent gauge coupling at the
Planck scale (denoted by $g$ in Sec.~\ref{sec:hidden}).\footnote{In the spirit
of string unified models one could assume that the observable and hidden gauge
couplings are related at the Planck scale or at the string scale ($M_{\rm
str}\approx4\times10^{17}\GeV$).} Fortunately, the issue of gauge coupling
unification in the observable sector has already been addressed in detail
in Ref.~\cite{HMTY}. Those calculations are directly applicable to our model
since the observable matter content and spectrum of the two models is the
same, even though the dynamics providing the intermediate scale are different.
Thus, here we limit ourselves to a brief summary of the relevant issues.

Writing down the one-loop RGEs for the gauge couplings, including a common
supersymmetric threshold at $M_{\rm SUSY}$, one can eliminate the unified
$SU(5)$ coupling and obtain the following two relations \cite{HMTY,Method}
\begin{eqnarray}
\left(3\alpha^{-1}_2-2\alpha^{-1}_3-\alpha^{-1}_1\right)(M_Z)&=&
{1\over2\pi}\left\{{12\over5}\ln{M_{H_3}M_{\bar H_3}\over M_{h'}M_Z}
-2\ln{M_{\rm SUSY}\over M_Z}-23.3\right\}
\label{eq:Relation1}\\
\left(5\alpha^{-1}_1-3\alpha^{-1}_2-2\alpha^{-1}_3\right)(M_Z)&=&
{1\over2\pi}\left\{36\ln{(M^2_V M_\Sigma)^{1/3}\over M_Z}
+8\ln{M_{\rm SUSY}\over M_Z}+12.1\right\}
\label{eq:Relation2}
\end{eqnarray}
In these relations, $M_V=3\sqrt{15}(g_5/\lambda_{75})M_{75}$ is the mass of the
GUT gauge bosons, and the explicit constants come from the splittings of the
\r{75} relative to the $M_\Sigma=5M_{75}$ mass of its (\r{8},\r{3}) component.
The above relations can be made more
accurate by the inclusion of realistic low-energy supersymmetric thresholds,
two-loop gauge coupling RGEs, and smooth decoupling of heavy particles. Once
this is done, and the latest values of the Standard Model gauge couplings
are input (\ie, $\alpha^{-1}=127.9\pm0.2$, $\sin^2\theta_W=0.2314\pm0.0004$,
$\alpha_3=0.118\pm0.007$), one obtains the following 1$\sigma$ allowed
intervals \cite{HMTY}:
\begin{eqnarray}
1.4\times10^{17}\GeV\le&{M_{H_3}M_{\bar H_3}\over M_{h'}}&
\le 5.5\times10^{20}\GeV\ ,
\label{eq:H3range}\\
8.4\times10^{15}\GeV\le&\left(M^2_V M_\Sigma\right)^{1/3}&
\le2.6\times10^{16}\GeV\ .
\label{eq:Vrange}
\end{eqnarray}
It is evident that our choices above, \ie, $M_{H_3}\sim M_{\bar H_3}\sim
10^{14}\GeV$ and $M_{h'}\sim10^{10}\GeV$, are perfectly consistent with the
constraint in Eq.~(\ref{eq:H3range}). The same is true for the
middle-of-the-road choice $M_V\sim M_\Sigma$ (\ie, $\lambda_{75}\sim g_5$),
which yields a GUT scale close to $10^{16}\GeV$.

We recall that we have set the masses of the \r{50},\rb{50} representations
at the gravitational scale $M\approx10^{18}\GeV$ in order to prevent the
onset of a non-perturbative $SU(5)$ regime below the Planck scale. Nonetheless,
because of the needed GUT-scale \r{75} representation, the unified gauge
coupling grows above the unification scale. However, it has been demonstrated
that this coupling remains in the perturbative regime, \ie, $\alpha\lsim0.1$
\cite{HMTY}. One could assume that the corresponding gauge coupling at the
gravitational scale ($g\approx1$) is related to the gauge coupling
from the hidden sector gauge group discussed in Sec.~\ref{sec:hidden}, as
would be the case in string models. This relation would help to further
constrain the viable hidden sector choices. For instance, assuming a
``super-unified" situation, where hidden and observable gauge couplings are
equal near the gravitational scale, the constraints on the hidden sector choice
can be read off Fig.~\ref{fig:NcNf} ($g=1$ curves).

Concerning proton decay, gauge-boson-mediated dimension-six operators depend
on $1/M^2_V$. From Eq.~(\ref{eq:Vrange}), $M_V$ is not expected to be much
below $10^{16}\GeV$, unless $\lambda_{75}\gg g_5$, but this case is unlikely
since $\lambda_{75}$ would be in the non-perturbative regime. Thus, we don't
expect a particular enhancement of dimension-six operators in this model. More
interesting is the situation with the dimension-five proton decay operators,
which depend on the effective Higgs triplet mass ($M_{H_{\rm eff}}$) defined
in Eq.~(\ref{eq:Heff}). The dominant proton partial lifetime is given by
\cite{HMY,HMTYpd}
\begin{equation}
\tau(p\to K^+\bar\nu_\mu)=2.0\times10^{31}\y
\left| {0.0056\GeV^3\over\beta}\,{0.67\over A_S}\,{\sin2\beta\over 1+y^{tK}}\,
{M_{H_{\rm eff}}\over 10^{17}\GeV}\,{\TeV^{-1}\over f}\right|^2\ ,
\label{eq:pdecay}
\end{equation}
where $\beta=(5.6\pm0.8)\times10^{-3}\GeV^3$ is the relevant hadronic matrix
element, $A_S$ is the short-distance renormalization factor, and $y^{tK}$
parametrizes the contribution of the third family relative to the first two
($|y^{tK}|\approx2$ for $m_t=175\GeV$) with an undetermined phase. The $f$
functions are the one-loop integrals which behave as $1/f\approx m^2_{\tilde
q}/m_{\widetilde W}$ for $m_{\tilde q}\gg m_{\widetilde W}$.

{}From unification constraints, Eq.~(\ref{eq:H3range}) indicates that
$M_{H_{\rm eff}}>1.4\times 10^{17}\GeV\approx10M_V$. In this case,
Eq.~(\ref{eq:pdecay})
and Ref.~\cite{AN} show that the present Kamiokande limit $\tau(p\to \bar\nu
K^+)>1\times10^{32}\y$ \cite{PDG}, is satisfied provided $\tan\beta$ is not too
large ($\tan\beta\lsim5$) and the universal scalar mass $m_0>300\GeV$. On the
other hand, in our model we obtain $M_{H_{\rm eff}}\gsim10^{18}\GeV
\approx100M_V$, and the experimental limit is satisfied rather comfortably,
even for large values of $\tan\beta$ and presently accessible supersymmetric
particle masses. For instance, for $m_{\tilde q}\approx 300\,(600)\GeV$ and
$m_{\widetilde W}\approx 80\GeV$, $\tan\beta\lsim10\,(40)$ is required. Thus,
$p\to \bar\nu K^+$  remains the dominant mode for proton decay, with good
prospects for observation at the upcoming SuperKamiokande experiment and the
proposed ICARUS facility. Note that the much-weakened proton-decay upper-bound
on $\tan\beta$ offers a new possibility in the study of Yukawa coupling
unification in $SU(5)$ GUTs (\ie, $\lambda_b=\lambda_\tau$), which now also
allow the so-called ``large-$\tan\beta$" solution \cite{YU}.

\section{Cosmic baryon asymmetry}
\label{sec:CBA}
With the realization of significant electroweak baryon number violation at high
temperatures, which occurs through ($B$+$L$)-violating but
($B$--$L$)-conserving non-perturbative sphaleron interactions
\cite{sphalerons}, several new mechanisms for generating the cosmic baryon
asymmetry have been proposed \cite{Olive}. These mechanisms produce a
primordial lepton asymmetry (leptogenesis), which is then recycled by
sphaleron interactions into a baryon asymmetry at the electroweak scale. It is
important to note that primordial ($B$--$L$)-conserving asymmetries, such as
those produced in traditional $SU(5)$ GUT baryogenesis, are likely to be wiped
out by the sphaleron interactions \cite{erasure}. Therefore, in the context of
$SU(5)$ GUTs, the leptogenesis-based mechanisms may be unavoidable. Here we
consider the simplest of these mechanisms, based on the out-of-equilibrium
decay of right-handed neutrinos, as first suggested in
Ref.~\cite{FY},\footnote{Before the realization of the importance of the
sphaleron interactions, Ref.~\cite{MNTY2} pointed out the possibility of
generating a baryon asymmetry in the decay of right-handed neutrinos via baryon
number violating GUT interactions.} and extended to supersymmetry in
Refs.~\cite{CDO,MSYY}, and to $SU(5)\times U(1)$ GUTs in Refs.~\cite{ENO,ELNO}.
We note that the lepton-asymmetric decays of right-handed sneutrino condensates
\cite{AD,sneutrinos}, may provide an additional contribution to the lepton
asymmetry that we discuss below.

In order to satisfy the out-of-equilibrium condition in the decay of the
right-handed neutrinos, one could follow the standard procedure of demanding
that the $\nu^c_{1,2,3}$ decay rate be less than the expansion rate of the
Universe at the time of $\nu^c$ decay. This condition leads to constraints on
the $\lambda_3$ couplings of the right-handed neutrinos, that can be
undesirable when trying to use the same couplings to compute the corresponding
light see-saw neutrino masses. Even more problematic can be the need to obtain
the surviving lepton asymmetry solely from the decays of the lightest
right-handed neutrino ($\nu^c_1$), since the asymmetry produced in the decays
of $\nu^c_{2,3}$ is typically wiped out by the $\nu^c_1$ interactions. Such
potential difficulties have been exemplified in Ref.~\cite{ELNO}. Instead, here
we follow an alternative scenario \cite{NOS},\footnote{Below we show that in
our model, the traditional out-of-equilibrium scenario is also viable.} whereby
the right-handed neutrinos are produced in the decays of the inflaton
subsequent to inflation. The COBE data on the anisotropy of the cosmic
microwave brackground radiation, interpreted in the context of inflation,
allows one to deduce the inflaton mass to be $m_{\eta}\sim 10^{11}\GeV$ and the
reheating temperature $T_R\sim10^8\GeV$ \cite{CDO}. The $\nu^c$ then decay
immediately after inflation and out of equilibrium at the temperature $T_R\ll
M_{\nu^c}$, {\em as long as} $M_{\nu_c}<m_{\eta}\sim10^{11}\GeV$.
Interestingly, the constraint from proton decay (see Sec.~\ref{sec:observable})
ensures that this condition is satisfied {\em automatically}.

The primordial lepton asymmetry, when reprocessed by sphaleron interactions,
leads to a similar baryon asymmetry \cite{CDO}
\begin{equation}
{n_B\over n_\gamma}\sim{n_L\over n_\gamma}\sim
\left({m_{\eta}\over M_{Pl}}\right)^{1/2}\epsilon\sim 10^{-4}\ \epsilon\ ,
\label{eq:nB}
\end{equation}
where the asymmetry parameter ($\epsilon$) due to the decay of the
$i$th-generation neutrino and sneutrino is given by \cite{CDO}
\begin{equation}
\epsilon_i = {1\over 2\pi (\lambda^\dagger_3\lambda_3)_{ii}}
\sum_j \left({\rm Im}\,[(\lambda^\dagger_3\lambda_3)_{ij}]^2\right)\,
g(M^2_{\nu^c_j}/M^2_{\nu^c_i})\ ,
\label{eq:eps}
\end{equation}
with
\begin{equation}
g(x)=4\sqrt{x}\,\ln{1+x\over x}\ .
\label{eq:g}
\end{equation}
To proceed we need to manipulate the entries in $\lambda_3$, which has
remained as yet unspecified. We define the unitary rotation matrix $U$,
such that $\hat\lambda_3=U\lambda_3 U^\dagger$, where $\hat\lambda_3$ is
the diagonal matrix of eigenvalues of $\lambda_3$. Experience with the
quark mixing matrix leads us to assume that $U$ differs little from the
identity matrix: $U={\bf1}+R$, with \cite{ELNO}
\begin{equation}
R=\left(\begin{array}{ccc} 0&\theta_{12}&0\\ -\theta^*_{12}&0&\theta_{23}\\
0&-\theta^*_{23}&0\end{array}\right)\ .
\label{eq:R}
\end{equation}
With this ansatz we obtain to lowest non-vanishing order
\begin{eqnarray}
(\lambda^\dagger_3\lambda_3)_{ii}&=&|\hat\lambda^i_3|^2
+\sum_j|\hat\lambda^j_3|^2\, |\theta_{ij}|^2 \ ,
\label{eq:ii}\\
(\lambda^\dagger_3\lambda_3)_{ij}&=&|\theta_{ij}|e^{i\phi_{ij}}
\left[|\hat\lambda^i_3|^2-|\hat\lambda^j_3|^2\right]\quad (i\not=j)\ ,
\label{eq:ij}
\end{eqnarray}
where $\phi_{ij}={\rm Arg}\,[\theta_{ij}]$. Thus, Eq.~(\ref{eq:eps}) becomes
\begin{equation}
\epsilon_i={1\over2\pi}\,
{\sum_{j\not=i}|\theta_{ij}|^2\sin2\phi_{ij}\,
[|\hat\lambda^i_3|^2-|\hat\lambda^j_3|^2]^2\
g(M^2_{\nu^c_j}/M^2_{\nu^c_i})\over
|\hat\lambda^i_3|^2+\sum_j|\hat\lambda^j_3|^2\,|\theta_{ij}|^2}\ .
\label{eq:eps2}
\end{equation}

Because of the several unknown parameters in the above expressions, and the
inherent uncertainties in this type of calculations, we will be content with
presenting a plausible scenario leading to interesting lepton asymmetries
and see-saw neutrino masses. For simplicity let us assume that the $\lambda_6$
matrix is proportional to the unit matrix, \ie,
\begin{equation}
M_{\nu^c_1}=M_{\nu^c_2}=M_{\nu^c_3}=M_{\nu^c}
=\lambda_6\,{\vev{T\bar T}\over M}\ .
\label{eq:Mnu^c}
\end{equation}
The light neutrino mass matrix then becomes $M_\nu=\lambda_3\lambda^T_3
v^2_2/M_{\nu^c}$. If we neglect the CP violating phases (a not necessarily
justified approximation), the matrix $U$ which diagonalizes
$\lambda_3\lambda^\dagger_3$, also diagonalizes $\lambda_3\lambda^T_3$ and the
physical neutrino masses become (up renormalization group scaling
corrections \cite{chorus})
\begin{equation}
m_{\nu_i}\approx{(\hat\lambda^i_3 v_2)^2\over M_{\nu^c}}\ .
\label{eq:Mnu}
\end{equation}
Furthermore, in our ansatz the (small) neutrino mixing angles are given by
$\theta_{e\mu}=\theta_{12}$, $\theta_{e\tau}=0$, and
$\theta_{\mu\tau}=\theta_{23}$. As we will see shortly, these mixing angles are
unrestricted from lepton asymmetry considerations, and thus could accomodate
the MSW solution to the solar neutrino problem ($\nu_e\leftrightarrow\nu_\mu$)
and lead to interesting $\nu_\mu\leftrightarrow\nu_\tau$ oscillations at the
CHORUS and NOMAD, and P803 experiments at CERN and Fermilab respectively.

{}From Eq.~(\ref{eq:Mnu}) we see that
$m_{\nu_\tau}\approx (\hat\lambda_3^3 v_2)^2/M_{\nu^c}
=[\hat\lambda_3^3\, \sin\beta(174\GeV)]^2/M_{\nu^c}$. With
$M_{\nu^c}\sim10^{10}\GeV$ and $\hat\lambda^3_3\approx0.1$, we get
$m_{\nu_\tau}\sim15\,(30)\eV$ for $\tan\beta\sim1\,(\tan\beta\gg1)$. This
range of tau neutrino masses provide an adequate and desirable hot dark matter
component of the Universe. Thus, in what follows we take $\hat\lambda^3_3=0.1$.
It is also natural to assume that the remaining eigenvalues of the $\lambda_3$
matrix are hierarchically smaller, \ie,
$\hat\lambda^1_3\ll\hat\lambda^2_3\ll\hat\lambda^3_3$. For instance,
$\hat\lambda^2_3\sim{1\over100} \hat\lambda^3_3$ yields $m_{\nu_\mu}\sim
10^{-3}\eV$, consistent with solutions to the solar neutrino problem via the
MSW mechanism. (These hierarchies are comparable to those in the up-quark
Yukawa matrix.)

Going back to the calculation of the lepton asymmetries, with our hierarchical
assumption for $\hat\lambda_3$, from Eq.~(\ref{eq:eps2}) we obtain
\begin{eqnarray}
\epsilon_1&\approx&{2\ln2\over\pi}\,\left(\hat\lambda^2_3\right)^2
\sin2\phi_{12}\sim 10^{-6}\,\phi_{12}\ ,
\label{eq:e1}\\
\epsilon_2&\approx&{2\ln2\over\pi}\,\left(\hat\lambda^3_3\right)^2
\sin2\phi_{23}\sim 10^{-2}\,\phi_{23}\ ,
\label{eq:e2}\\
\epsilon_3&\approx&{2\ln2\over\pi}\,\left(\hat\lambda^3_3\right)^2
|\theta_{23}|^2\sin2\phi_{23}\sim  10^{-2}\,|\theta_{23}|^2\phi_{23}\ .
\label{eq:e3}
\end{eqnarray}
With the expression for the estimated baryon asymmetry in Eq.~(\ref{eq:nB}), we
would get the desired result of ${\rm few}\times10^{-10}$ for $\phi_{12}\sim1$
and $\phi_{23}\ll1$. The natural choice would be maximal CP violation in
the $\theta_{12}$ entry in the rotation matrix $R$ (see Eq.~(\ref{eq:R}))
and no CP violation elsewhere in the matrix (unless new entropy diluting
sources are introduced to reduce $\epsilon_1+\epsilon_2+\epsilon_3$). These
results would be affected somewhat if one allows a non-trivial structure
to the matrix $\lambda_6$ (\ie, relaxing the assumption in
Eq.~(\ref{eq:Mnu^c})).

We now remark that this model is also viable in the traditional
out-of-equilibrium scenario, where $\epsilon_1$ is the only surviving
asymmetry. The out-of-equilibrium condition at $T=M_{\nu^c_1}\sim10^{10}\GeV$,
\begin{equation}
\Gamma_{\nu^c_1}={(\lambda^\dagger_3\lambda_3)_{11}\over16\pi}\,
 M_{\nu^c_1}<1.66 g^{1/2}_*\,{T^2\over M_{Pl}}=H\ ,
\label{eq:out}
\end{equation}
is satisfied for (using Eq.~(\ref{eq:ii}))
\begin{equation}
(\lambda^\dagger_3\lambda_3)_{11}=
|\hat\lambda^1_3|^2+|\hat\lambda^2_3|^2\,|\theta_{12}|^2\lsim10^{-6}\ ,
\label{eq:cond}
\end{equation}
which is consistent with our hierarchical assumption. However, in this case
the calculation of the leptonic asymmetry has a larger ($\sim10^{-2}$
\cite{CDO}) coefficient than in Eq.~(\ref{eq:nB}), requiring a non-maximal CP
violating phase $\phi_{12}\sim10^{-2}$.

Finally, let us comment on whether or not the sphaleron interactions may wash
away the leptonic asymmetry produced above. This could in principle happen if
the non-renormalizable operators obtained when integrating out the right-handed
neutrino fields, \ie, $(\lambda_3/M_{\nu^c}) LLHH$, where $L$ is the lepton
doublet in $\bar f$ and $H$ the Higgs doublet in $h$, are in equilibrium with
the sphaleron interactions \cite{FY}. It has been shown \cite{CKO} that to
prevent the erasure of the asymmetry, one must demand
$M_{\nu^c}\gsim(\lambda_3)^2\, 3\times10^9\GeV$, which is always satisfied for
our choices of $\lambda_3$ and $M_{\nu^c}$.

\section{Comparison with $SU(5)\times U(1)$}
\label{sec:comparison}
The revamped MDM presented in the previous sections has several appealing
phenomenological features, constituting an interesting example of traditional
grand unified model building. Nonetheless, it is apparent that the model
is rather non-minimal or uneconomical. For instance, a \r{75} needs to be
used for GUT symmetry breaking, greatly increasing the size of the GUT particle
spectrum. Moreover, the \r{50},\rb{50} to effect the doublet-triplet splitting
problem make the unified gauge coupling so large above the GUT scale that they
need to be taken at the gravitational scale. The doublet-triplet splitting
is tamed, but proton decay can still be too fast because of the ``useless"
pieces of the \r{50},\rb{50} representations which need to be made heavy,
resulting in the otherwise-not-needed doubling of these representations and of
the Higgs pentaplets. Regarding the right-handed neutrinos, their (ad-hoc)
introduction has various desirable consequences. However, the Yukawa matrix
coupling them to the lepton doublets is arbitrary, with no particular
motivation for its desired hierarchical structure.

It is interesting to note that the above critique of the revamped MDM can be
circumvented altogether if one extends the gauge group from $SU(5)$ to
$SU(5)\times U(1)$ \cite{Barr,revitalized,Moscow}. Gauge symmetry breaking
down to the Standard Model gauge group occurs via vacuum expectation values
of the $H$ (\r{10}) and $\bar H$ (\rb{10}) Higgs representations. This is
possible because of the ``flipping" $u\lra d$, $u^c\lra d^c$, $e\lra\nu$,
$e^c\lra\nu^c$ involved in the assignment of the Standard Model particles
to the $\bar f=\{u^c,L\}$ (\rb{5}) and $F=\{Q,d^c,\nu^c\}$ (\r{10})
representations. Thus, $H$ and $\bar H$ contain one pair of neutral fields
$\nu^c_H,\nu^c_{\bar H}$, which get vevs along the flat direction
$\vev{\nu^c_H}=\vev{\nu^c_{\bar H}}$. There is no need for large GUT
representations for symmetry breaking. As is well known (and we review below),
this property takes on a much larger magnitude when one attempts to derive
these models in string model building.

The missing-partner mechanism, which above involved the couplings
$\bar\theta\Sigma h$ [(\rb{50})(\r{75})(\r{5})] and $\theta\Sigma\bar h$
[(\r{50})(\r{75})(\rb{5})], is now effected by the couplings $HHh$
[(\r{10})(\r{10})(\r{5})] and $\bar H\bar H\bar h$
[(\rb{10})(\rb{10})(\rb{5})]. First note that no additional representations
are needed besides the GUT-breaking Higgs ones. Moreover, the resulting
Higgs triplet matrix
\begin{equation}
\bordermatrix{
&\bar h_3&d^c_H\cr
h_3&0&\lambda_4\,\vev{\nu^c_H}\cr
d^c_{\bar H}&\lambda_5\,\vev{\nu^c_{\bar H}}&0}\ ,
\label{eq:f2/3}
\end{equation}
does not need a large non-zero (22) entry ({\em c.f.} Eq.~(\ref{eq:2/3}))
because the ``useless" components of the $H$ and $\bar H$ representations
are eaten by the GUT gauge bosons to become massive or become GUT Higgsinos.
This natural zero mass term for $d^c_H d^c_{\bar H}$ implies that the
dimension-five proton decay operators are negligible. We end up with a very
economical GUT Higgs spectrum and no threat of dimension-five operators.

Regarding neutrino masses, the right-handed neutrinos which had to be
introduced by hand in the revamped MDM, are now contained in the $F$ (\r{10})
representations. Indeed, the coupling $\lambda_3\bar f\nu^c h$ in
Eq.~(\ref{eq:W'''}) is here written as $\lambda_3\bar f e^c h$, with the
(unavoidable) right-handed electrons now introduced ``by hand". In $SU(5)\times
U(1)$ this coupling provides the charged lepton masses. On the other hand, the
coupling $\lambda_1 F\bar f\,\bar h$, which in Eq.~(\ref{eq:W}) provided the
down-quark masses, here provides the up-quark masses and Dirac neutrino masses.
(Also, the coupling $\lambda_2 FFh$, which in Eq.~(\ref{eq:W}) provided the
up-quark masses, here provides the down-quark masses.) Thus, the right-handed
neutrinos are unavoidable in $SU(5)\times U(1)$, and their Yukawa couplings
to the lepton doublets are equal to those of the up-quark Yukawa matrix,
providing (as discussed in Sec.~\ref{sec:CBA}) an automatic and desirable
hierarchy in the see-saw neutrino masses. An important distinction between
the see-saw mechanism in the revamped MDM and $SU(5)\times U(1)$ is the
manner in which the right-handed neutrinos get a mass. In the revamped MDM
this is through the superpotential term $\lambda_6\nu^c\nu^c\vev{T\bar T}/M$ in
Eq.~(\ref{eq:W'''}), whereas in $SU(5)\times U(1)$ there are two possible
sources: (i) through cubic couplings
$\lambda_6 F\bar H\phi\ni\lambda_6\vev{\nu^c_{\bar H}}\nu^c\phi$, where $\phi$
(with $\vev{\phi}=0$) are $SU(5)$ singlets \cite{revitalized}; and (ii)
through non-renormalizable couplings $\lambda_9 FF\bar H\bar H/M\ni
\lambda_9 (\vev{\nu^c_{\bar H}}^2/M)\nu^c\nu^c$ \cite{improved}. The second
form resembles that in the revamped MDM, although the mass scale is likely
higher (\ie, $\vev{\nu^c_{\bar H}}^2/M\sim10^{14}\GeV$).

These two models also differ somewhat in the calculation of the cosmic baryon
asymmetry, besides the possible difference in the right-handed neutrino mass
spectrum. Indeed, because the $SU(5)\times U(1)$ gauge symmetry is broken along
a flat direction, there is a dilution factor ($\Delta$) in the computation of
the lepton asymmetry due to the entropy released by the late-decaying ``flaton"
field \cite{dilution,ENO}. However, these two effects ($\nu^c$ spectrum and
$\Delta$) tend to compensate each other and an acceptable baryon asymmetry is
typically obtained \cite{ENO,ELNO}.

There is another cosmological aspect of these models that sets them apart,
namely the breaking of the GUT symmetry down to the Standard Model one.
In the MDM, $SU(5)$ symmetry breaking via an arbitrary vev of the \r{75} leads
to several possible degenerate vacua \cite{HMPR}, at least in the context of
global supersymmetry. When supergravity effects are taken into account, if the
desired vacuum has zero cosmological constant, all the others will be lower in
energy, although essentially unreachable \cite{Weinberg}. Thus, if the vev
of the \r{75} can be arranged to be in the desired direction, the Universe
will remain in the desired broken phase. In contrast, in $SU(5)\times U(1)$
the breaking down to the Standard Model via the vevs of the \r{10},\rb{10}
along the F- and D-flat direction $\vev{\nu^c_H}=\vev{\nu^c_{\bar H}}$
is {\em unique} \cite{revitalized}.

Regarding the issue of unification, the revamped MDM requires non-minimal
representations to make this possible. In $SU(5)\times U(1)$ traditional
grand unification does not occur (although the non-abelian Standard Model gauge
groups do unify) and unification is not a test of the model. However, if string
unification is desired (at the scale $M_{\rm str}\approx4\times10^{17}\GeV$),
then non-minimal representations need to be added to the $SU(5)\times U(1)$
model \cite{gap}.

We have seen that the pair of \r{50},\rb{50} representations in the revamped
MDM need to be put at the gravitational scale. It is then natural to ask
whether this model can be obtained from the only known consistent theory of
quantum gravity, namely string theory. Because of some technical difficulties
which we review below, no attempts have been made to derive the MDM from
strings. It is of course well known that $SU(5)\times U(1)$ can be easily
derived from strings \cite{revamp,search}.

The prime constraint in string model-building is that of the massless
representations which are allowed when the corresponding gauge group $G$ is
represented by a ``level-$k$" Kac-Moody algebra on the world-sheet
\cite{ELN,GO}. The allowed representations must be unitary,
\begin{equation}
\sum_{i=1}^{\rm rank\, G} n_i m_i\le k\ ,
\label{eq:unitary}
\end{equation}
where $n_i$ are the Dynkin labels of the highest weight representation in
question, and $m_i$ are fixed positive integers for a given $G$. In the case of
$SU(n)$: $m_i=1\,,\,\forall i$. In our $SU(5)$ example then $\sum_{i=1}^4
n_i\le k$. Looking up the $n_i$ values, we see that for $k=1$, only
\r{1},\r{5},\rb{5},\r{10},\rb{10} are unitary. For $k=2$ we find in addition:
\r{15},\r{24},\r{40},\rb{40},\r{45},\rb{45},\r{50},\rb{50},\r{75}. Only
level-one constructions appear to be needed to derive $SU(5)\times U(1)$,
whereas at least level-two constructions are required in the MDM. However,
this is not the end of the story, since one can also ask whether the allowed
representations could possible be massless. This requires calculating the
so-called conformal dimension $h_r$ of the representation $r$,
\begin{equation}
h_r={C_r\over 2k+C_A}\ ,
\label{eq:h}
\end{equation}
where $C_r$ is the Casimir of $r$, and $C_A$ that of the adjoint
representation. If $h_r>1$, the representation is necessarily massive. For
$h_r\le1$ the representation could be massless, although this is not guaranteed
since other degrees of freedom may add their own contribution to the conformal
dimension making it exceed unity. It is not hard to see that in $SU(5)$ all
unitary representations at level one are also massless \cite{ELN}, and thus
$SU(5)\times U(1)$ models can be readily constructed at level one. The unitary
representations of interest for MDM model-building, which are allowed at level
two, have conformal dimensions
\begin{equation}
h_{50,\overline{50}}={42\over5(k+5)}\ ,\qquad h_{75}={8\over k+5}\ ,
\label{eq:h5075}
\end{equation}
and are not massless at level two. In fact, $k=4$ is required to make all these
representations massless. Such high-level Kac-Moody constructions have never
been attempted.

One intriguing possibility would be to construct level-two $SU(5)$ string
models (for recent attempts see Ref.~\cite{Aldazabal}), which should allow the
required large MDM representations, although with masses at the Planck scale.
Note that this is not necessarily a problem since we already require the
\r{50},\rb{50} to be at that mass scale. If the \r{75} is also raised to that
scale, the breaking of $SU(5)$ would occur at the string scale, and this may be
difficult to reconcile with gauge coupling unification. It is also worth
remarking that in a string model all gauge couplings are related at the string
scale, and with $SU(5)$ constructed at level two, the relation would be
$\sqrt{2}\,g_5=g_h$ \cite{Ginsparg}, where $g_h$ is the gauge coupling of the
hidden gauge group. Finally, the mass terms in Eqs.~(\ref{eq:W},\ref{eq:W'}),
which would not be allowed if the large MDM representations belonged to the
massless spectrum, are expected to arise when they belong to the string massive
spectrum. Of course, bridging the gap between the massless and massive spectrum
may create problems in obtaining the low-energy effective field theory, but
this question cannot be answered until an actual model is constructed along
these lines.

\section{Conclusions}
\label{sec:conclusions}
During the last few years, a great deal of attention has been paid to
supersymmetric grand unified theories in light of the precise LEP measurements
of the Standard Model gauge couplings. These analyses depend crucially on
the details of the low-energy supersymmetric spectrum and the heavy GUT
spectrum. Most of the effort to date has been focused on the minimal $SU(5)$
supergravity model, which appears to be running into difficulties regarding
unification and proton decay. In addition, there is the nagging doublet-triplet
splitting problem that receives no satisfactory explanation. Motivated by
these developments, we have reconsidered one of the alternatives to minimal
$SU(5)$, where the doublet-triplet splitting is dealt with in a reasonable
way via the missing-partner mechanism, and gauge coupling unification is not
in jeopardy. We have revamped this model to tame the dimension-five proton
decay operators, and to allow see-saw neutrino masses. In order to generate the
needed intermediate scale for the right-handed neutrino masses, we have endowed
the model with a ``modern" hidden sector which can generate dynamically the
desired intermediate scale, the scale of supersymmetry breaking, and the Higgs
mixing parameter $\mu$. The revamped MDM also provides for the cosmic baryon
asymmetry through the Fukugita-Yanagida mechanism via lepton-number-violating
decays of the right-handed neutrinos.

We have also contrasted the main features of the revamped MDM against the
``flipped" $SU(5)\times U(1)$ model, and basically shown that the former
can be considered as  a ``poor man's" version of the latter. In the realm
of string model-building, $SU(5)\times U(1)$ fares rather well, whereas
the revamped MDM is very unlikely to be realized, except perhaps if one
allows $SU(5)$ symmetry breaking to occur at the string scale.

\section*{Acknowledgments}
We would like to thank John Ellis and Bruce Campbell for useful discussions.
This work has been supported in part by DOE grant DE-FG05-91-ER-40633.

\newpage

\clearpage


\begin{thebibliography}{99}
\bibitem{EKN}J. Ellis, S. Kelley, and \DVN, \PLB{249}{90}{441} and
\PLB{260}{91}{131}; P. Langacker and M. Luo, \PRD{44}{91}{817}; U. Amaldi, W.
de Boer, and H. F\"urstenau, \PLB{260}{91}{447}; F. Anselmo, L. Cifarelli, A.
Peterman, and A. Zichichi, Nuovo Cim. {\bf104A} (1991) 1817. For a review see,
\eg, \JL, \DVN, and \AZ, Prog. Part. Nucl. Phys. {\bf 33} (1994) 303.
\bibitem{GUTthresholds}J. Ellis, S. Kelley and D. V.  Nanopoulos,
\NPB{373}{92}{55}; R. Barbieri and L. Hall, \PRL{68}{92}{752}; J. Hisano, H.
Murayama, and T. Yanagida, \PRL{69}{92}{1014}; K. Hagiwara and Y. Yamada,
\PRL{70}{93}{709}; F. Anselmo, L. Cifarelli, A. Peterman, and A. Zichichi,
Nuovo Cim. {\bf105A} (1992) 1025; P. Langacker and N. Polonsky,
\PRD{47}{93}{4028}.
\bibitem{CAN}A. Chamseddine, R. Arnowitt, and P. Nath, \PRL{49}{82}{970}.
\bibitem{smooth} M. Bastero-Gil and J. Perez-Mercader, \PLB{322}{94}{355}
and LAEFF-95-01 \hepph{9506222}; A. Faraggi and B. Grinstein, \NPB{422}{94}{3};
P. Chankowski, Z. Pluciennik, and S. Pokorski, \NPB{439}{95}{23}.
\bibitem{Clavelli} L. Clavelli and P. Coulter, \PRD{51}{95}{3913} and
UAHEP954 \hepph{9507261}.
\bibitem{Bagger}J. Bagger, K. Matchev, and D. Pierce, \PLB{348}{95}{443}.
\bibitem{LP} P. Langacker and N. Polonsky, UPR-0642T \hepph{9503214}.
\bibitem{NROs}D. Ring, S. Urano, and R. Arnowitt, CTP-TAMU-01/95
\hepph{9501247}; T. Dasgupta, P. Mamales, and P. Nath \hepph{9501325}.
\bibitem{DG}S. Dimopoulos and H. Georgi, \NPB{193}{81}{150}; N. Sakai, Z. Phys.
C{\bf11} (1981) 153.
\bibitem{MPM} A. Masiero, \DVN, K. Tamvakis, and T. Yanagida,
\PLB{115}{82}{380}; B. Grinstein, \NPB{206}{82}{387}.
\bibitem{SSM}E. Witten, \PLB{105}{81}{267}; \DVN\ and K. Tamvakis,
\PLB{113}{82}{151}; L. Iba\~nez and G. Ross, \PLB{110}{82}{215}.
\bibitem{PGBM}K. Inoue, A. Kakuto, and T. Takano, Prog. Theor. Phys. {\bf75}
(1986) 664; A. Anselm and A. Johansen, \PLB{200}{88}{331}; R. Barbieri, G.
Dvali, and A. Strumia, \NPB{391}{93}{487}; R. Barbieri, G. Dvali, and M.
Moretti, \PLB{312}{93}{137}; R. Barbieri, \etal, \NPB{432}{94}{491};
Z. Berezhiani, C. Csaki, and L. Randall, \NPB{444}{95}{61}; Z. Berezhiani,
INFN-FE 02-95 \hepph{9503366}.
\bibitem{Lahanas} J. Polchinski and L. Susskind, \PRD{26}{82}{3661}; H. Nilles,
M. Srednicki, and D. Wyler, \PLB{124}{82}{237}; A. Lahanas, \PLB{124}{82}{341}.
\bibitem{MNTY2} A. Masiero, \DVN, K. Tamvakis, and T. Yanagida, Z. Phys. C
{\bf17} (1983) 33.
\bibitem{HMTY}J. Hisano, T. Moroi, K. Tobe, and T. Yanagida,
\PLB{342}{95}{138}.
\bibitem{AN}R. Arnowitt and P. Nath, \PRL{69}{92}{725}; P. Nath and
R. Arnowitt, \PLB{287}{92}{89}; R. Arnowitt and P. Nath, \PRD{49}{94}{1479}.
\bibitem{HMY}J. Hisano, H. Murayama, and T. Yanagida, \NPB{402}{93}{46}.
\bibitem{LNP}\JL, \DVN, and A. Zichichi, \PLB{291}{92}{255};
\JL, \DVN, and H. Pois, \PRD{47}{93}{2468}; R. Arnowitt and P. Nath,
\PLB{299}{93}{58} and {\bf307} (1993) 403(E); P. Nath and R. Arnowitt,
\PRL{70}{93}{3696}; \JL, \DVN, and K. Yuan, \PRD{48}{93}{2766}.
\bibitem{HMTYpd}J. Hisano, T. Moroi, K. Tobe, and T. Yanagida, TU-470
\hepph{9411298}.
\bibitem{GFM}H. Georgi, AIP Conference Proceedings No.23, New York 1975, p.
575; H. Fritzsch and P. Minkowski, Ann.~Phys.~{\bf93} (1975) 193.
\bibitem{GN}H. Georgi and \DVN, \NPB{155}{79}{52} and \NPB{159}{79}{16}.
\bibitem{BB} K. Babu and S. Barr, \PRD{48}{93}{5354}, \PRD{50}{94}{3529},
BA-95-11 \hepph{9503215}, BA-95-21 \hepph{9506261}.
\bibitem{Mohapatra}D. Lee and R. Mohapatra, \PLB{324}{94}{376} and
\PRD{51}{95}{1353}
\bibitem{Raby}G. Anderson, S. Dimopoulos, L. Hall, S. Raby, and G. Starkman,
\PRD{49}{94}{3660}; L. Hall and S. Raby, \PRD{51}{95}{6524}.
\bibitem{inspired}K. Babu and S. Barr, \PRD{51}{95}{2463}; K. Babu and
R. Mohapatra, \PRL{74}{95}{2418}; D. Lee and R. Mohapatra, UMD-PP-95-93
\hepph{9502210}; B. Brahmachari and R. Mohapatra, UMD-PP-95-138
\hepph{9505347}.
\bibitem{NR}A. Nelson and L. Randall, \PLB{316}{93}{516}.
\bibitem{Carena}M. Carena, S. Pokorski, M. Olechowski, and C. Wagner,
\NPB{426}{94}{269}; R. Ratazzi, U. Sarid, and L. Hall, \PRD{50}{94}{7048}.
\bibitem{BOP}F. Borzumati, M. Olechowski, and S. Pokorski, \PLB{349}{95}{311}.
\bibitem{OP}M. Olechowski and S. Pokorski, \PLB{344}{95}{201}.
\bibitem{EG+NS}J. Ellis and M. Gaillard, \PLB{88}{79}{315}; \DVN\ and M.
Srednicki, \PLB{124}{83}{37}.
\bibitem{Lykken}S. Chaudhuri, S. Chung, and J. Lykken,
Fermilab-Pub-94-137-T\\ \hepph{9405374}; G. Cleaver, OHSTPY-HEP-T-94-007
\hepth{9409096}.
\bibitem{Cleaver}S. Chaudhuri, S. Chung, G. Hockney, and J. Lykken,
Fermilab-Pub-94-413-T \hepph{9501361}; G. Cleaver, OHSTPY-HEP-T-95-003
\hepth{9505080} and OHSTPY-HEP-T-95-004 \hepth{9506006}.
\bibitem{ELN}J. Ellis, \JL, and \DVN, \PLB{245}{90}{375}; A. Font, L. Iba\~nez,
and F. Quevedo, \NPB{345}{90}{389}.
\bibitem{Aldazabal}G. Aldazabal, A. Font, L. Iba\~nez, and A. Uranga,
FTUAM-28/94 \hepph{9410206}; G. Aldazabal, \hepth{9507162}.
\bibitem{MSY}H. Murayama, H. Suzuki, and T. Yanagida, \PLB{291}{92}{418}.
\bibitem{decisive}See \eg, J. L. Lopez and \DVN, \PLB{251}{90}{73},
\PLB{256}{91}{150}, and \PLB{268}{91}{359}.
\bibitem{CKN}E. Chun, J. Kim, and H. Nilles, \NPB{370}{92}{105}; J. Kim and
H. Nilles, \MODA{9}{94}{3575}.
\bibitem{Casasmu}J. Casas and C. Mu\~noz, \PLB{306}{93}{288}.
\bibitem{Peskin}N. Evans, S. Hsu, and M. Schwetz, YCTP-P8-95 \hepth{9503186};
O. Aharony, J. Sonnenschein, M. Peskin, and S. Yankielowicz, SLAC-PUB-6938
\hepth{9507013}.
\bibitem{Seiberg} See \eg, N. Seiberg, \PRD{49}{94}{6857} and references
therein.
\bibitem{Method} See also, J. Hisano, H. Murayama, and T. Yanagida in
Ref.~\cite{GUTthresholds} and Ref.~\cite{HMTYpd}.
\bibitem{PDG}Particle Data Group, \PRD{50}{94}{1173}.
\bibitem{YU}S. Kelley, \JL, and \DVN, \PLB{274}{92}{387}; V. Barger, M. Berger,
and P. Ohmann, \PRD{47}{93}{1093}; G. Anderson, S. Dimopoulos, L. Hall, and S.
Raby, \PRD{47}{93}{3702}.
\bibitem{sphalerons} V. Kuzmin, V. Rubakov, and M. Shaposhnikov,
\PLB{155}{85}{36}. For a review see, A. Cohen, D. Kaplan, and A. Nelson,
Ann.~Rev.~Nucl.~Part.~Sci. {\bf43} (1993) 27.
\bibitem{Olive}For a recent review see, K. Olive, UMN-TH-1319/94
\hepph{9503342}.
\bibitem{erasure} P.~Arnold and L.~McLerran, \PRD{36}{87}{581} and
\PRD{37}{88}{1020}.
\bibitem{FY}M. Fukugita and T. Yanagida, \PLB{174}{86}{45}. See also
M. Fukugita and T. Yanagida, \PRD{42}{90}{1285}; M. Luty, \PRD{45}{92}{455}.
\bibitem{CDO}B. Campbell, S. Davidson, and K. Olive, \NPB{399}{93}{111}.
\bibitem{MSYY} H. Murayama, H. Suzuki, T. Yanagida, J. Yokohama,
\PRL{70}{93}{1912} and \PRD{50}{94}{2356}.
\bibitem{ENO}J. Ellis, \DVN, and K. Olive, \PLB{300}{93}{121}.
\bibitem{ELNO}J. Ellis, \JL, \DVN, and K. Olive, \PLB{308}{93}{70}.
\bibitem{AD}I. Affleck and M. Dine, \NPB{249}{85}{361}. For recent work see, M.
Dine, L. Randall, and S. Thomas, \PRL{75}{95}{395} and SLAC-PUB-95-6846
\hepph{9507453}.
\bibitem{sneutrinos}B. Campbell, S. Davidson, and K. Olive, \PLB{303}{93}{63};
H. Murayama and T. Yanagida, \PLB{332}{94}{349}.
\bibitem{NOS}A. Dolgov and A. Linde, \PLB{116}{82}{329}; \DVN, K. Olive, and
M. Srednicki, \PLB{127}{83}{30}.
\bibitem{chorus}J. Ellis, \JL, and \DVN, \PLB{292}{92}{189}.
\bibitem{CKO}J. Cline, K. Kainulainen, and K. Olive, \PRL{71}{93}{2372} and
\PRD{49}{94}{6394}.
\bibitem{Barr} S. Barr, \PLB{112}{82}{219}, \PRD{40}{89}{2457}; J. Derendinger,
J. Kim, and \DVN, \PLB{139}{84}{170}.
\bibitem{revitalized}I. Antoniadis, J. Ellis, J. Hagelin, and \DVN,
\PLB{194}{87}{231}.
\bibitem{Moscow} For reviews see \DVN, in ``Les Rencontres de Physique de
la Vallee d'Aoste: Proceedings: Supernova 1987A,  one year later: results and
perspectives in particle physics." Edited by M. Greco (Editions Frontieres,
1988), p. 795; \JL, Surveys~H.~E.~Phys. {\bf8} (1995) 135 \hepph{9405278}.
\bibitem{improved}I. Antoniadis, J. Ellis, J. Hagelin, and \DVN,
\PLB{208}{88}{209}.
\bibitem{dilution}B. Campbell, J. Ellis, J. Hagelin, \DVN, and K. Olive,
\PLB{197}{87}{355}.
\bibitem{HMPR}T. H\"ubsch, S. Meljanac, S. Pallua, and G. Ross,
\PLB{161}{85}{122}.
\bibitem{Weinberg}S. Weinberg, \PRL{48}{82}{1776}.
\bibitem{gap}I. Antoniadis, J. Ellis, S. Kelley, and \DVN, \PLB{272}{91}{31};
S. Kelley, \JL, and \DVN, \PLB{278}{92}{140};
D. Bailin and A. Love, \PLB{280}{92}{26}; G. Leontaris, \PLB{281}{92}{54};
\JL, \DVN, and A. Zichichi, \PRD{49}{94}{343}.
\bibitem{revamp}I. Antoniadis, J. Ellis, J. Hagelin, and \DVN,
\PLB{231}{89}{65}.
\bibitem{search}\JL, \DVN, and K. Yuan, \NPB{399}{93}{654}.
\bibitem{GO}For a review see, P. Goddard and D. Olive, \IJMP{1}{86}{303}.
\bibitem{Ginsparg}P. Ginsparg, \PLB{197}{87}{139}.
\end{thebibliography}
\end{document}